\documentclass{emulateapj}
\usepackage{amsmath}

\slugcomment{Accepted for publication in ApJ}
\shorttitle{Time Delays in H1413+117}
\shortauthors{L.J. Goicoechea \& V.N. Shalyapin}

\begin{document}

\title{Time Delays in the Gravitationally Lensed Quasar H1413+117 (Cloverleaf)}

\author{Luis J. Goicoechea}
\affil{Departamento de F\'\i sica Moderna, Universidad de Cantabria, Avda. de Los Castros s/n, 
39005 Santander, Spain}
\email{goicol@unican.es}

\and

\author{Vyacheslav N. Shalyapin}
\affil{Institute for Radiophysics and Electronics, National Academy of Sciences of Ukraine, 12 
Proskura St., 61085 Kharkov, Ukraine}
\email{vshal@ukr.net}

\begin{abstract}
The quadruple quasar H1413+117 ($z_s$ = 2.56) has been monitored with the 2.0 m Liverpool Telescope 
in the $r$ Sloan band from 2008 February to July. This optical follow--up leads to accurate light 
curves of the four quasar images (A--D), which are defined by 33 epochs of observation and an 
average photometric error of $\sim$ 15 mmag. We then use the observed (intrinsic) variations of 
$\sim$ 50$-$100 mmag to measure the three time delays for the lens system for the first time 
(1$\sigma$ confidence intervals): $\Delta \tau_{AB}$ = $-$17 $\pm$ 3, $\Delta \tau_{AC}$ = $-$20 
$\pm$ 4, and $\Delta \tau_{AD}$ = 23 $\pm$ 4 days ($\Delta \tau_{ij} = \tau_j - \tau_i$; B and C 
are leading, while D is trailing). Although time delays for lens systems are often used to obtain 
the Hubble constant ($H_0$), the unavailability of the spectroscopic lens redshift ($z_l$) in the 
system H1413+117 prevents a determination of $H_0$ from the measured delays. In this paper, the new 
time delay constraints and a concordance expansion rate ($H_0$ = 70 km s$^{-1}$ Mpc$^{-1}$) allow 
us to improve the lens model and to estimate the previously unknown $z_l$. Our 1$\sigma$ estimate 
$z_l$ = 1.88$^{+0.09}_{-0.11}$ is an example of how to infer the redshift of very distant galaxies 
via gravitational lensing. 
\end{abstract}


\section{Introduction}          
The time delay between two images of a gravitationally lensed source depends on the distribution of
mass in the lens and the current expansion rate of the Universe \citep{refsdal64,refsdal66}. This 
expansion rate is quantified by the Hubble constant $H_0$. If the source is a quasar, the intrinsic 
quasar variability may be used to determine the time delays between its multiple images. Thus, 
observed delays for lensed quasars lead to valuable information on $H_0$, provided lensing mass 
distributions can be constrained by observational data 
\citep[e.g.,][]{kochanek04,schechter04,saha06,jackson07,oguri07}. Future large samples of lens 
systems could be useful tools to obtain accurate estimates of the main cosmological parameters 
\citep[e.g.,][]{dobke09}. 

For a given lensed quasar, each time delay between two of its images is indeed scaled by a factor 
containing $H_0$, the lens redshift $z_l$, and additional physical parameters. If $z_l$ is known 
(this is the usual situation), the Hubble constant and the lensing mass distribution can be 
simultaneously deduced by using a parametric lens scenario and a set of observational constraints 
\citep[including information on the time delay(s); e.g.,][]{schneider06}. However, a large set of 
constraints is required to reliably determine both cosmological and galactic properties. 

Accurate observations of a quadruply imaged quasar in a well--studied galaxy field bring an 
excellent opportunity to study in detail $H_0$ and the mass distribution of the gravitational lens. 
Apart from the three independent time delays, and the positions and fluxes of the four images, data 
on the neighbour galaxies also constraint the lens scenario \citep[e.g.,][]{jackson07}. 
Alternatively, if $z_l$ is very uncertain or unknown, one may infer a lens redshift value from the 
time delay measurements and the rest of constraints (using a value of $H_0$ from other experiments). 

In this paper we focus on the quadruple quasar \object{H1413+117} \citep[Cloverleaf;][]{magain88}, 
lying at a redshift $z_s$ = 2.558 \citep[e.g.,][]{barbainis97}.
Although the lens redshift is currently unknown, several neighbour galaxies were detected by 
\citet{kneib98}. \citet{kneib98} found the main lensing galaxy G1 amid the four quasar images, and 
presented astrometric and photometric data for additional galaxies surrounding \object{H1413+117}. 
They analysed a secondary lensing galaxy (G2) close to G1, as well as some other objects probably 
belonging to an overdensity (galaxy group/cluster) at photometric redshift $z_{ove} \sim$ 0.9. 
\citet{faure04} also found evidence for the presence of two different overdensities at $z_{ove}$ = 
0.8 $\pm$ 0.3 \citep[corresponding to the structure discovered by][]{kneib98} and $z_{ove}$ = 1.75 
$\pm$ 0.2, which could contribute noticeably to the lensing potential.  

Very recently, \citet{macleod09} have used the positions of G1, G2, and other candidate lensing 
galaxies, the quasar image positions \citep{turnshek97}, new mid--IR flux ratios, and some priors 
to constrain the lensing mass distribution. They have concluded that the galaxy pair G1--G2 and an 
external shear (likely related to the observed galaxy overdensities) are required to explain the 
observations. From an optical monitoring over the 1987--94 period, \citet{ostensen97} also reported 
a quasi--simultaneous variability of the four quasar images. Unfortunately, the scarce sampling did 
not permit them to estimate the time delays in the system. Hence, the measurement of time delays 
for the Cloverleaf quasar should improve knowledge about the lens (mass and redshift), as well as 
allow completion of new studies (e.g., microlensing variability). 

In Section 2 we present new optical light curves of \object{H1413+117}. In Section 3, from these 
light curves, we estimate the time delays between quasar images. In Section 4 we compare the most
recent lens scenario with all relevant observations. In Section 5 we discuss our results and put
them into perspective.

\section{Observations and light curves}
We observed \object{H1413+117} from early February to late July of 2008, i.e., for 6 months. All
observations were made with the 2.0 m fully robotic Liverpool Telescope (LT) at the Roque de los 
Muchachos Observatory, Canary Islands (Spain), using the RATCam optical CCD camera (binning 2 
$\times$ 2). The global database consists of 61 exposures of 300 s in the $r$ Sloan 
filter\footnote{The pre--processed frames are publicly available on the Liverpool Quasar Lens
Monitoring archive at \url{http://dc.zah.uni--heidelberg.de/liverpool/res/rawframes/q/form}.}. 
These original exposures (frames) are almost regularly distributed over the whole observation 
period (there is only a significant 21--day gap in 2008 February), with an average sampling rate 
of about one frame each three days. In Fig. 1, a combined LT frame shows the blended quasar images 
(at the centre of the field), two relevant stars \citep[the control star S40 and the PSF star S45, 
which correspond to the objects 40 and 45 in Fig. 1 of][]{kayser90}, and other relatively bright 
objects. The left bottom corner of Fig. 1 illustrates the positions (crosses) and names (A--D) of 
the four quasar images.

The four quasar images are separated by $\sim 1\arcsec$ (see Fig. 1), so we only consider 33 
high--quality frames to make the light curves (photometry on the other 28 frames is discussed in 
the last paragraph of Section 3). This selection is based on the FWHM of the seeing disc measured 
in each frame (FWHM $< 1\farcs 5$), as well as the signal--to--noise ratio (SNR) of the 18.16--mag 
control star (SNR $>$ 150). We note that this star (S40 in Fig. 1) has an $r$--band brightness 
similar to those of the quasar images ($\sim$ 17.9--18.4 mag). Each SNR value is calculated within 
an aperture with radius equal to the frame FWHM.

We determine the instrumental fluxes of the four quasar images through point--spread function 
(PSF) fitting. As the main lensing galaxy is very faint in the $r$ band \citep[$R >$ 22.7 
mag;][]{kneib98}, our 
photometric model includes four stellar--like sources (i.e., 4 empirical PSFs) plus a constant 
background. The empirical PSF is derived from the 16.69--mag star in the vicinity of the lens 
system (S45 in Fig. 1). In order to obtain accurate and reliable fluxes, we use the well--tested 
IMFITFITS software \citep{macleod98}, incorporating the relative positions of the B--D images 
\citep[with respect to A;][]{turnshek97} as constraints. Thus, the code is applied to all selected 
frames (see above), allowing 7 parameters to be free. These free parameters are the position of 
the A image, the fluxes of the A--D images, and the sky brightness. We also infer SDSS magnitudes 
from the relative instrumental magnitudes of the lensed quasar. The S45 star is taken as reference 
for differential photometry. In Table~\ref{tbl1} we present the $r$--SDSS magnitudes of the four 
images and the control (S40) star. 

From the standard deviation of the magnitudes of the S40 star over the whole monitoring period, 
we obtain a typical error of 0.006 mag. This global scatter agrees with the standard deviation 
between magnitudes on consecutive nights divided by the square root of 2, as expected on 
theoretical grounds. To estimate typical photometric errors in the quasar light curves, we then 
use the standard deviations between magnitudes having time separations $<$ 1.5 days (true 
variability is negligible on this very short timescale), which are divided by the square root of 2. 
The resulting uncertainties are 0.010 (A), 0.012 (B), 0.018 (C) and 0.018 (D) mag. As a summary, 
we achieve $\sim$ 1--2\% photometry and reasonable sampling rate ($\sim$ 6 data per month). 
Moreover, the light curves of the A--D images show significant variations of $\sim$ 0.05--0.1 mag 
(see Fig. 2). For example, the almost parallel fading by $\sim$ 0.1 mag of A--D (over the last 100 
days in Fig. 2) suggests intrinsic variability. This is promising to derive time delays. 

\section{Time delays}
There are four different image ray paths for the Cloverleaf quasar, so traveltime ($\tau$) varies 
from image to image \citep{schneider92}. Assuming that the observed magnitude fluctuations are 
basically originated in the source quasar (intrinsic variability), we use two well--known 
cross--correlation techniques to measure time delays between quasar images $\Delta \tau_{ij} = 
\tau_j - \tau_i$, where $i,j = A, B, C, D$. These techniques are the dispersion ($D^2$) and 
reduced chi--square ($\hat{\chi}^2$) minimizations \citep[e.g.,][]{pelt96,ullan06}. Despite the 
existence of other methods for determining time delays \citep[e.g.,][]{kundic97,gil02}, most 
methods work in a similar way, and in most cases one does not need to carry out an exhaustive 
analysis. After deriving delays we discuss the intrinsic variability hypothesis at the end of this 
section.

We focus on the AB, AC, and AD comparisons, i.e., the A light curve is compared to the other 
three brightness records (B--D). The $D^2$ and $\hat{\chi}^2$ minimizations are characterized by a 
decorrelation length ($\delta$) and a bin semisize ($\alpha$), respectively. To simultaneously 
avoid very noisy trends and loss of signal (due to excessive smoothing), we take $\alpha$ = 
$\delta$ = 15 days. For $\alpha$ = $\delta$ = 15 days, the spectra between $-$75 and +75 days 
include global and local minima. However, these local minima do not play an important role in 
the estimation of the time delays $\Delta \tau_{AB}$, $\Delta \tau_{AC}$, and $\Delta \tau_{AD}$
(see below).  

For a given cross--correlation method, we follow two different approaches to generate synthetic 
light curves and determine time delay errors. In the first approach (which is called NORMAL), we do 
not make any hypothesis on the underlying signal, but the observational noises (in the four light 
curves) are assumed to be normally distributed. Therefore, one obtains a synthetic light curve 
of an image by adding a random quantity to each brightness in the observed record. These random 
quantities are realizations of a normal distribution around zero, with a standard deviation equal 
to the standard deviation between observed magnitudes on consecutive nights. We produce 1000 
synthetic light curves of each image, and thus, obtain 1000 delay values for each pair (AB, AC, and 
AD) and the corresponding 68\% confidence intervals. In the second approach, we use a bootstrap 
procedure \citep[BOOTSTRAP; e.g.,][]{efron93}. First, we derive a combined light curve from the 
best solutions of $\Delta \tau_{AB}$, $\Delta \tau_{AC}$, and $\Delta \tau_{AD}$ (global minima of
the three spectra), and the associated magnitude differences. This combined curve is then smoothed 
by a 20--day filter. Second, the combined and smoothed curve is assumed to be a rough reconstruction 
of the underlying signal, and thus, the residuals for each image are taken as sets of errors. These
four sets are resampled to infer 1000 bootstrap simulations (synthetic light curves) of each image. 
To measure the time delays (68\% confidence intervals), we compute 1000 delay values for each pair 
of images and analyse the three distributions. 

The time delay measurements are presented in Table~\ref{tbl2}. For each pair of images, the four 
measurements are consistent with each other. Table~\ref{tbl2} also indicates the existence of an 
average offset of about 3 days between the results from the NORMAL approach and those derived from 
BOOTSTRAP simulations. This offset may be due to slightly biased reconstructions of the underlying 
signal (BOOTSTRAP procedure). We adopt $\Delta \tau_{AB}$ = $-$17 $\pm$ 3, $\Delta \tau_{AC}$ = 
$-$20 $\pm$ 4, and $\Delta \tau_{AD}$ = 23 $\pm$ 4 days as our final 1$\sigma$ measurements 
($\hat{\chi}^2$ and NORMAL). The corresponding $r$--band magnitude differences are $\Delta m_{AB}$ = 
0.155 $\pm$ 0.006, $\Delta m_{AC}$ = 0.322 $\pm$ 0.011, and $\Delta m_{AD}$ = 0.501 $\pm$ 0.007 mag 
(1$\sigma$ confidence intervals). These give the magnification (flux) ratios: $B/A$ = 0.867 $\pm$ 
0.005, $C/A$ = 0.743 $\pm$ 0.007, and $D/A$ = 0.630 $\pm$ 0.004.

From the central values in the time delay and magnitude difference intervals, one can 
obtain a final combined light curve, i.e., the A light curve together with the magnitude-- and 
time--shifted B--D records. In Fig. 3 we display this combined light curve (filled symbols). Using a 
20--day filter (filter semisize = 10 days), a possible reconstruction of the underlying intrinsic 
signal is also drawn in Fig. 3 (solid line). The standard deviation between the magnitudes in the 
combined record and the reconstruction is about 0.015 mag (see the shaded area in Fig. 3, which 
represents 0.015--mag deviations from the reconstruction), in good agreement with the average 
photometric error in the quasar light curves (see the error bar in the lower left corner of Fig. 3). 
This suggests the absence of significant microlensing signals, and strengthens our hypothesis that 
observed variations mainly have an intrinsic origin. We also note that \citet{ostensen97} reported 
on an almost parallel variation in brightness ($R$ band) of the four quasar images over a 7--year 
period (see the middle panels in Figs. 2--3 of that paper). Moreover, our 6--month monitoring period 
was long enough to find significant variability and determine time delays, but it was not long 
enough to detect substantial microlensing variations. Typical microlensing gradients of $\sim$ 
10$^{-4}$ mag day$^{-1}$ are expected in the $rR$ bands 
\citep[e.g.,][]{gaynullina05,fohlmeis07,shalyapin09}.   

Once the time delays are measured and the combined light curve is drawn in Fig. 3 (filled symbols), 
we can discuss the accuracy of the magnitudes derived from the poor--quality frames (see Section 2). 
These 28 exposures with FWHM $\geq 1\farcs 5$ and/or SNR $\leq$ 150 seem to be useless, since they 
lead to a very noisy combined record and confusion. One frame has a very poor image quality, so we
do not extract quasar magnitudes. Some of the magnitudes associated with the other 27 poor--quality 
frames are shown in Fig. 3 (open symbols). The rest are extremely noisy, and their values are outside 
the magnitude range in Fig. 3. In brief, 28 out of the 61 original frames have either an excessive 
blurring, or an insufficient signal, or both of them, so they do not produce accurate quasar light 
curves.

\section{Improved lens model and lens redshift}
In a model--independent way, the image and main lens positions for the \object{H1413+117} system are 
useful to determine the ordering of the time delays \citep{saha03}. However, the quasar images B and 
C (associated with minimum arrival times) are almost equidistant from the main lens, so it is 
difficult to distinguish the leading image in this system. In any case, intrinsic variations should 
be firstly observed in light curves of these two images, and later in records of A and D. While D is 
the trailing image (it is the closest to the main lens), A should be characterized by an intermediate 
arrival time. Our time delay measurements agree with this time--ordering of the images. Detailed lens 
models predict that C is leading \citep{chae99,macleod09}. However, we cannot confirm this prediction 
at 1$\sigma$ confidence level, since the $\Delta \tau_{AB}$ and $\Delta \tau_{AC}$ intervals overlap 
each other (a direct measure $\Delta \tau_{BC}$ is neither useful to decide on the leading image).

\citet{macleod09} reported how a relatively simple lens model reproduces the observed positions and 
mid--IR flux ratios of the four quasar images. These mid--IR flux ratios are insensitive to 
extinction (long wavelength) and microlensing (large emission region). The MacLeod et al.'s main 
solution (see the second column in Table~\ref{tbl3}) relies on an observationally motivated scenario. 
This consists of a background point--like source (quasar) that is lensed by a singular isothermal 
ellipsoid (main lensing galaxy G1), a singular isothermal sphere (secondary lensing galaxy G2), and 
an external shear \citep[likely produced by galaxy overdensities;][]{kneib98,faure04}. Although the 
position of G1 was constrained by observations, it was allowed to vary during the fitting procedure. 
The singular isothermal sphere was placed at the observed position of G2, and MacLeod et al. also 
assumed priors on the ellipticity of G1 ($e_{G1}$ = 0.0 $\pm$ 0.5) and the strength of the external 
shear ($\gamma_{ext}$ = 0.05 $\pm$ 0.05). 
 
Here, we use the MacLeod et al.'s lens scenario and a concordance cosmology: $H_0$ = 70 km s$^{-1}$ 
Mpc$^{-1}$, $\Omega_m$ = 0.3, $\Omega_{\Lambda}$ = 0.7 \citep[e.g.,][]{spergel03}. The theoretical 
time delay between two lensed images includes a cosmological scale factor $T = (1 + z_l)(D_l D_s/ 
cD_{ls})$, where $c$ is the velocity of light and $D$ denotes angular diameter distance 
\citep[e.g.,][]{schneider92}. The angular diameter distances are determined by the cosmology, and the 
lens and source redshifts, $z_l$ and $z_s$. Thus, using a concordance cosmology and the observed 
redshift of the source $z_s$ = 2.558 \citep[e.g.,][]{barbainis97}, the scale factor exclusively 
depends on $z_l$. Our goal is to find a good fit to all observations of interest, i.e., the image 
positions and (mid--IR) fluxes, and the three time delays in Section 3. We also use the constraints 
and priors on the G1--G2 positions, $e_{G1}$, and $\gamma_{ext}$ by \citet{macleod09}. With respect 
to the MacLeod et al.'s framework, we add 3 new observational constraints (time delays) and one new 
free parameter ($z_l$), so the degrees of freedom (dof) change from 5 to 7. 

Through the GRAVLENS\footnote{\url{http://redfive.rutgers.edu/$\sim$keeton/gravlens}.} software 
package \citep{keeton01}, we find a solution with $\chi^2$/dof = 7.5/7. Our main results are shown in 
the third column of Table~\ref{tbl3}. Most lensing mass parameters of this acceptable solution are 
close (deviations less than 10\%) to those in the second column of the same table \citep[see also the 
third column in Table 3 of][]{macleod09}. However, it is interesting to note that the new external 
shear strength ($\gamma_{ext}$ = 0.11) is slightly larger than the previous one, whereas the mass 
scale of G2 ($b_{G2} = 0\farcs 45$) is slightly smaller (deviations of 25--30\%). On the other hand, 
the best--fit lens redshift indicates the existence of a very distant galaxy pair G1--G2. In Fig. 4 we 
draw the $\chi^2-z_l$ relationship, which permits us to obtain confidence intervals for $z_l$. The 
1$\sigma$ determination is $z_l$ = 1.88$^{+0.09}_{-0.11}$ (dark shaded area in Fig. 4). Fig. 4 also 
shows the 2$\sigma$ (95\%) confidence interval: 1.65 $\leq z_l \leq$ 2.03 (whole shaded area). 
   
\section{Discussion} 
In the lens system \object{H1413+117}, the main lensing galaxy G1 is a very faint object \citep[$R >$ 
22.7 mag;][]{kneib98}, surrounded by four close and relatively bright quasar images ($R \sim$ 18 mag). 
Thus, it is very difficult to separate the spectrum of the galaxy from those of the quasar images. 
Moreover, the quasar spectra indicate the presence of intervening objects (absorption lines) at 
different redshifts less than $z_s$ = 2.56 \citep[e.g.,][and references therein]{monier98}, so one 
cannot decide on the redshift of G1. The secondary lensing galaxy G2 and other galaxies in the 
vicinity of the lensed quasar are also very faint objects \citep[$R >$ 23 mag;][]{kneib98}, and
spectroscopic redshifts of these galaxies are not available yet. Photometric data of field galaxies
are consistent with the presence of two galaxy overdensities at $z_{ove}$ = 0.8 $\pm$ 0.3 and 
$z_{ove}$ = 1.75 $\pm$ 0.2 \citep{kneib98,faure04}. For example, the galaxy G2 has a photometric 
redshift of about 2 \citep{kneib98}. Our gravitational lensing estimate of the redshift of G1--G2: 
$z_l$ = 1.88$^{+0.09}_{-0.11}$ (1$\sigma$ interval), is in reasonable agreement with the photometric 
redshift of G2 and the most distant overdensity, as well as the absorption system at $z_{abs}$ = 1.87. 
However, the nearest group/cluster is far away from the principal gravitational deflector (galaxy pair 
G1--G2). 

So far it is not considered a possible uniform external convergence, i.e., $\kappa_{ext}$ = 0. If 
$\kappa_{ext} \neq$ 0, then fits to $\kappa_{ext}$ = 0 lens scenarios (lens models with 
$\kappa_{ext}$ = 0) must be conveniently rescaled. Original estimates of $H_0$ (if that were the case)
or $z_l$ (in our case) also require suitable corrections. This is because of the so--called mass sheet 
degeneracy \citep[e.g.,][]{falco85,gorenstein88,saha00,nakajima09}. For \object{H1413+117}, it is 
unclear what is the main perturber producing external shear, and perhaps external convergence. For
example, infrared photometry of the neighbour galaxy H2 is consistent with a redshift of about 2 
\citep{kneib98}, so it could be located in the principal lens plane. Moreover, the orientation of the 
external shear, $\theta_{\gamma_{ext}} = 45\fdg 4$, is in the direction of this galaxy \citep[e.g., 
see Fig. 4 of][]{macleod09}. Thus, H2 and some other related objects (belonging to the very distant 
overdensity) might produce most of the external shear and a negligible convergence. Apart from this 
optimistic perspective, one may also consider that the main perturber is the most distant group/cluster 
as a whole. If this were true, the external convergence at the quasar positions would be $\kappa_{ext}$ 
= $\gamma_{ext} \sim$ 0.1 for a singular isothermal sphere. Taking $\kappa_{ext} \sim$ 0.1 into 
account, the derived lens redshift increases by $\sim$ 3\% (+ 0.05). This increase represents only one 
half of our 1$\sigma$ uncertainty in $z_l$ ($\sim$ 0.10; see above).
 
We can also quantify the gravitational influence of the nearest overdensity. A weak lensing analysis 
of this structure indicated a shear direction of $45\degr$ \citep[see the last column in Table 5 
of][]{faure04}. This shear direction coincides with our strong lensing determination in 
Table~\ref{tbl3}, which suggests that the nearest overdensity might play a noticeble role in the lens
phenomenon. \citet{faure04} also determined an upper limit on the shear strength: $\gamma <$ 0.17 
\citep[see Fig. 7 of][]{faure04}. \citet{keeton03} and \citet{momcheva06} discussed the effective 
convergence and shear when a perturber does not lie at $z_l$, e.g., $z_{per} < z_l$. Assuming a 
singular isothermal sphere to describe the perturber, $\kappa_{eff} = \gamma_{eff} \sim (1 - \beta)
\gamma$, where $\beta \sim$ 0 if $z_{per} \sim z_l$ and $\beta >$ 0 if $z_{per} < z_l$ \citep[see 
Appendix of][]{momcheva06}. From the involved redshifts and the upper limit on the shear strength, we 
infer that $\kappa_{eff} = \gamma_{eff} <$ 0.03. Thus, the nearest structure cannot account for the 
external shear in the lens system, and it likely generates no more than 10\% of the total shear 
strength. The constraint on the effective convergence leads to a negligible increase in $z_l$, i.e., 
$\Delta z_l$ is below + 0.01. In some other lens systems, there are also perturbers at redshifts less 
than those of the principal lensing objects, which produce a few hundredths of external convergence 
and shear \citep[e.g.,][]{fassnacht02,fassnacht06}. 

Our lens scenario incorporates singular isothermal mass profiles. Despite the fact that these profiles
lead to an acceptable fit ($\chi^2$/dof = 7.5/7), other distributions of mass (including a core, some 
deviation from the isothermal behaviour or both ingredients) could also lead to good fits for the 
observational data. This is the well--known profile degeneracy \citep[e.g.,][]{jackson07}. An 
exhaustive study of mass distributions consistent with the available observational constraints is out 
of the scope of this paper. However, we check the influence of non--isothermal profiles of G1 on $z_l$ 
estimations. The power--law index of G1 ($\alpha$) is assumed to be around 1 (isothermal index), and 
additional fits with singular $\alpha \neq 1$ profiles are done. For $\alpha$ = 1.1, the solution is 
characterized by $\Delta \chi^2 = \chi^2 - \chi^2(\alpha = 1)$ = 0.1. For $\alpha$ = 0.8--0.9, we find 
a very modest improvement in $\chi^2$, $\Delta \chi^2$ = $-$0.1. The full range $\alpha$ = 0.8--1.1 
leads to best--fit values of $z_l$ within our 1$\sigma$ "isothermal" estimate in Section 4. 

Microlensing variations in \object{H1413+117} may shed light on the nature and structure of the source 
quasar \citep[e.g.,][]{lewis98,popovic06}. The three time delay measurements in Section 3 and the new 
data on the lens (mass and redshift) in Section 4 are useful tools for analyses of microlensing 
variability. Once the delays are known, it is possible to properly compare quasar light curves and to 
search for microlensing signals. Moreover, the (lens and source) redshifts and the improved lens model 
allow construction of microlensing magnification patterns and simulated microlensing light curves 
\citep[e.g.,][]{wambsganss90}. The key parameters for microlensing simulations are the convergence and 
shear strength at the positions of the quasar images. To address the space distribution of both 
convergence and shear, we consider our lens model in the third column of Table~\ref{tbl3}. This gives
($\kappa_A$, $\gamma_A$) = (0.51, 0.58), ($\kappa_B$, $\gamma_B$) = (0.52, 0.32), ($\kappa_C$, 
$\gamma_C$) = (0.48, 0.37), and ($\kappa_D$, $\gamma_D$) = (0.57, 0.65). Although we cannot rule out 
the existence of an external convergence at a level of 0.1, this scenario is uncertain (see above). 
Hence, we do not take into account any external convergence due to galaxy overdensities along the line 
of sight to the lensed quasar.

Our lens model causes different magnifications at the four positions of the quasar images. The model
magnification ratios are $B/A$ = 0.82, $C/A$ = 0.78, and $D/A$ = 0.45. All these ratios agree, as 
expected, with the mid--IR measurements by \citet{macleod09}. However, the model ratios are not 
included in the error bars of our optical ($r$--band) flux ratios in Section 3. These differences 
between optical and model ratios are probably due to dust extinction and microlensing magnification.
If microlensing is currently playing a role \citep[e.g., the optical continuum of the image D could be 
magnified by microlensing;][and references therein]{anguita08}, it should be a long--term effect that 
induces small (optical) flux variations on time scales of several months. This microlensing 
variability scenario is supported by recent studies for other systems with non--local lens galaxies 
\citep[e.g.,][]{gaynullina05,fohlmeis07,shalyapin09}.  

\acknowledgments

The Liverpool Telescope is operated on the island of La Palma by Liverpool John Moores University in 
the Spanish Observatorio del Roque de los Muchachos of the Instituto de Astrofisica de Canarias with 
financial support from the UK Science and Technology Facilities Council. We thank the Liverpool 
Telescope staff for kind interaction over the observation period (program ID: CL08A03). We use 
datasets taken from the Sloan Digital Sky Survey (SDSS) Web site, and we are grateful to the SDSS team
for doing that public database. This research has been supported by the Spanish Department of 
Education and Science grant AYA2007-67342-C03-02, and University of Cantabria funds. 

{\it Facilities:} \facility{Liverpool:2m (RATCam)}.

\clearpage

\begin{figure}
\plotone{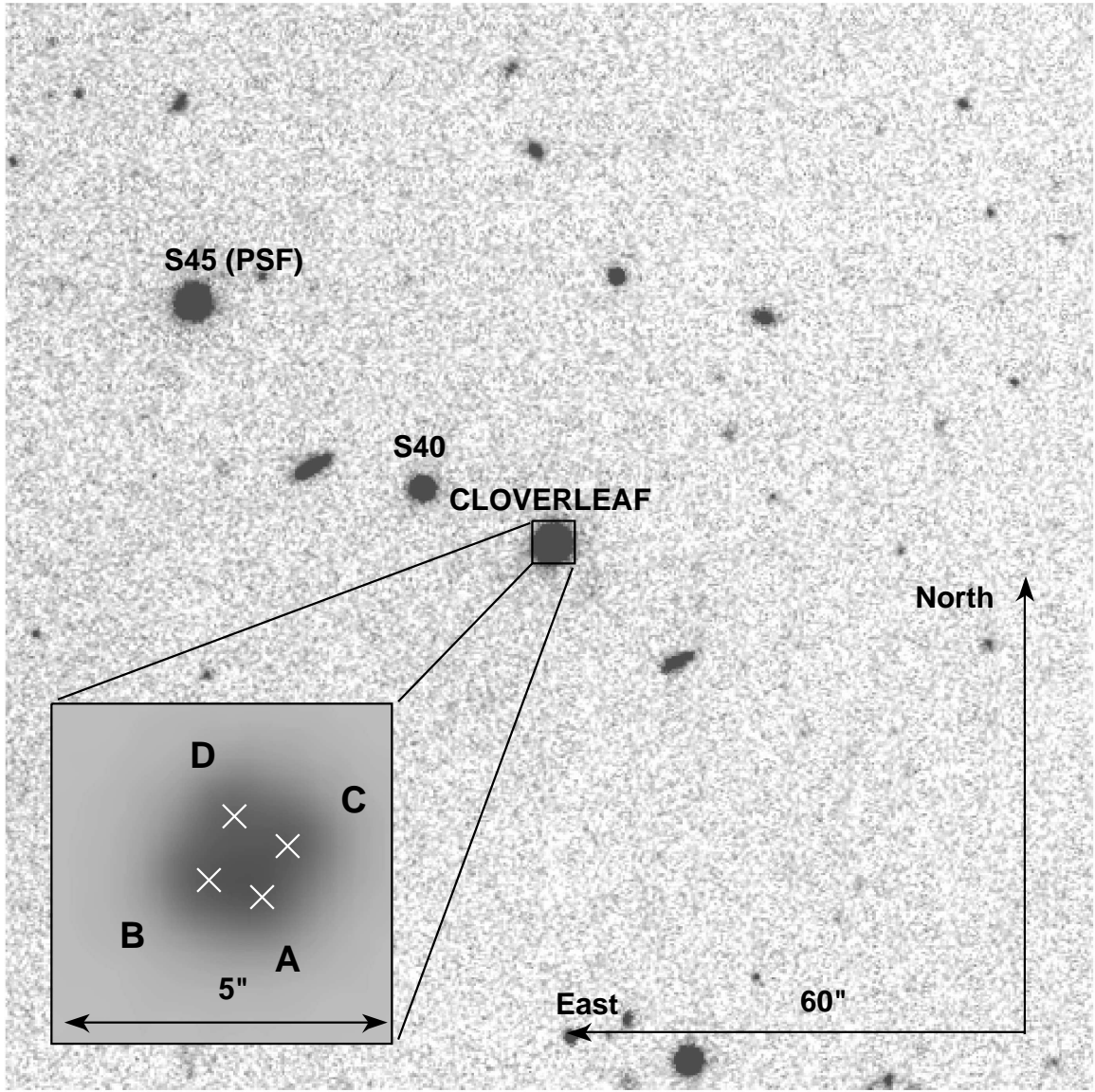}
\caption{$r$--band Liverpool Telescope imaging of the Cloverleaf quasar. We combine the 5 best 
frames in terms of seeing (total exposure time = 1500 s, FWHM = $0\farcs 87$). We then zoom into 
the central region and perform linear interpolation between adjacent pixels (box in the left 
bottom corner). The four quasar images and two relevant stars are properly labeled.}
\end{figure}

\clearpage

\begin{figure}
\plotone{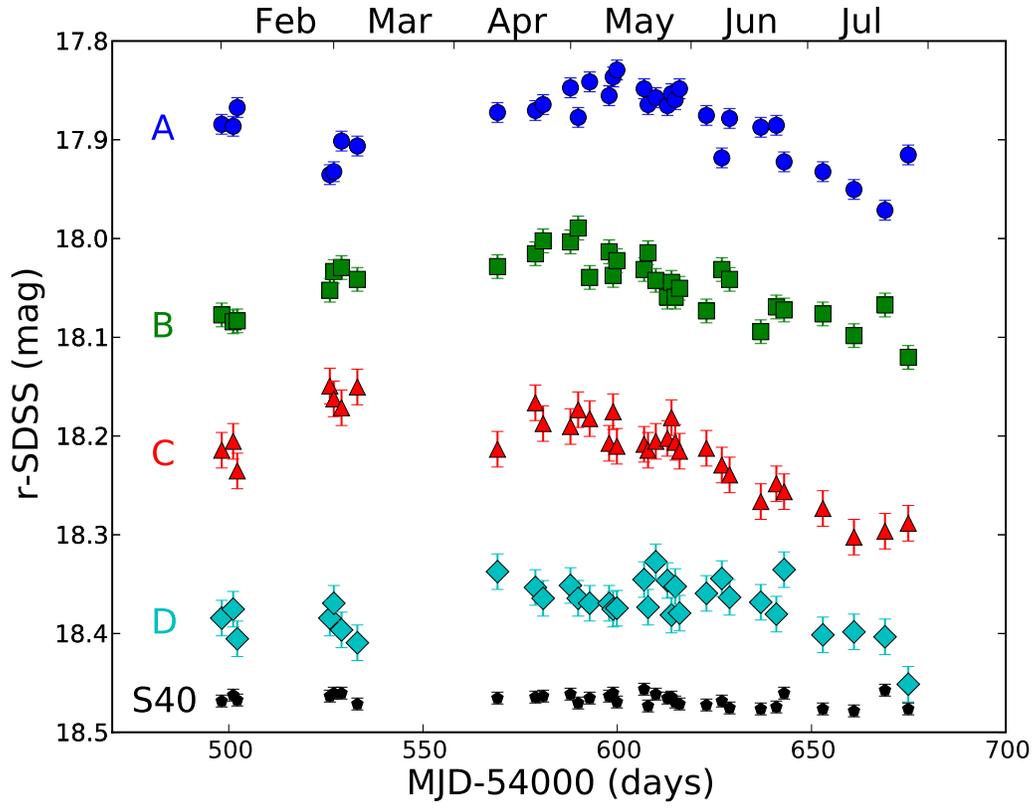}
\caption{Light curves of four quasar images A--D and the control star S40. The stellar record is 
shifted by +0.3 mag to facilitate comparison.} 
\end{figure}

\clearpage

\begin{figure}
\plotone{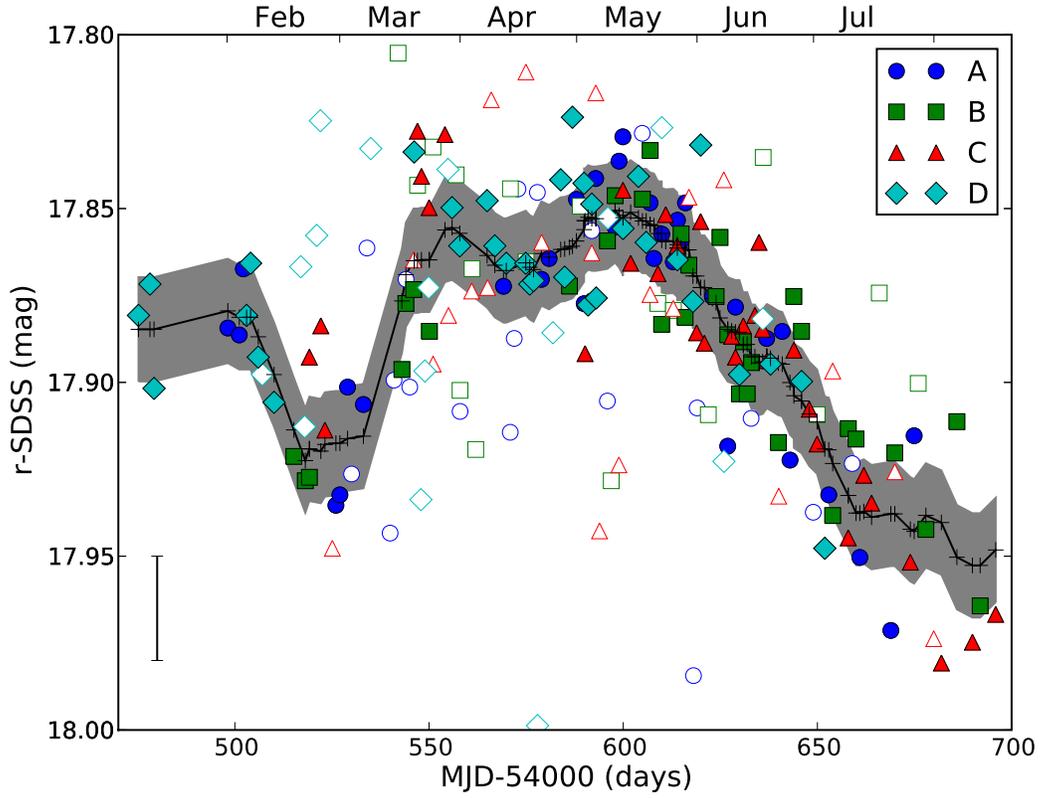}
\caption{Final combined light curve of H1413+117. This is made from the A light curve 
and the magnitude-- and time--shifted B--D records (filled symbols). To create the combined 
record, we use the time delays $\Delta \tau_{AB}$ = $-$17, $\Delta \tau_{AC}$ = $-$20, and 
$\Delta \tau_{AD}$ = 23 days (shifts in time), and the magnitude differences $\Delta m_{AB}$ = 
0.155, $\Delta m_{AC}$ = 0.322, and $\Delta m_{AD}$ = 0.501 mag (shifts in magnitude). We also 
display a possible reconstruction of the quasar signal (solid line), as well as the standard 
deviation between the combined record and the reconstruction (shaded area). The error bar in 
the lower left corner represents the average photometric error in the quasar light curves. To 
check the accuracy of the quasar magnitudes in 27 poor--quality frames that are not used in 
the time delay analysis, we also compare some of these additional magnitudes (open symbols) 
with those derived from the high--quality exposures (filled symbols). Several additional data 
(for the poor--quality frames) are outside the magnitude range from 18.0 to 17.8.} 
\end{figure}

\clearpage

\begin{figure}
\plotone{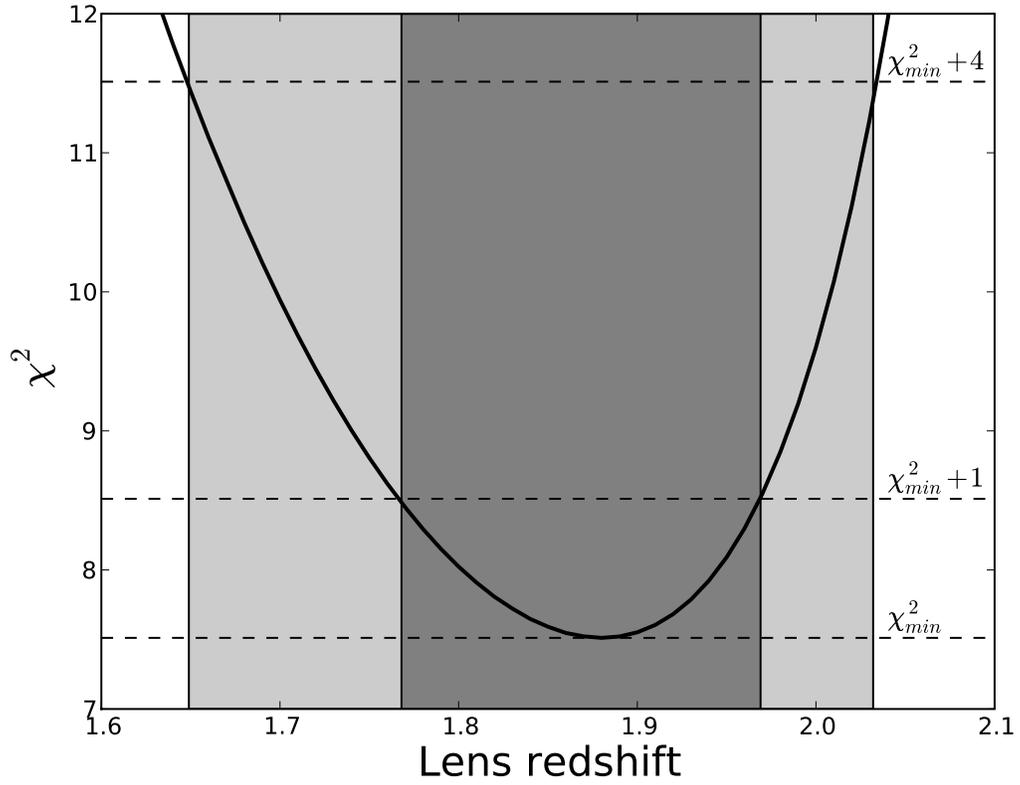}
\caption{Estimation of the previously unknown lens redshift. We show both the 1$\sigma$ (dark 
shaded area) and 2$\sigma$ (whole shaded area) confidence intervals. We estimate $z_l$ via 
gravitational lensing, using observational constraints on the lensed quasar images (positions, 
fluxes, and time delays), as well as data on the neighbour galaxies.} 
\end{figure}

\clearpage

\begin{deluxetable}{cccccccccc}
\tablecaption{Photometry of H1413+117.\label{tbl1}}
\tablewidth{0pt}
\tablehead{
\colhead{number} & \colhead{civil date\tablenotemark{a}} & \colhead{MJD-54000} & 
\colhead{FWHM\tablenotemark{b}} & 
\colhead{SNR\tablenotemark{c}} & 
\colhead{A\tablenotemark{d}} & \colhead{B\tablenotemark{e}} & \colhead{C\tablenotemark{f}} & 
\colhead{D\tablenotemark{g}} & \colhead{S40\tablenotemark{h}}
}
\startdata
1  & Feb 1   &  498.1885  &  1.08  &  235  &  17.884  &  18.077  &  18.214  &  18.384  &  18.168 \\
2  & Feb 4   &  501.1370  &  1.07  &  232  &  17.886  &  18.084  &  18.205  &  18.375  &  18.162 \\
3  & Feb 5   &  502.1893  &  1.26  &  233  &  17.867  &  18.083  &  18.235  &  18.405  &  18.167 \\
4  & Feb 29  &  526.0175  &  1.45  &  164  &  17.935  &  18.052  &  18.149  &  18.384  &  18.163 \\
5  & Mar 1   &  527.0400  &  1.24  &  186  &  17.932  &  18.033  &  18.162  &  18.369  &  18.160 \\
6  & Mar 3   &  529.0227  &  1.44  &  173  &  17.901  &  18.029  &  18.171  &  18.396  &  18.160 \\
7  & Mar 7   &  533.0942  &  1.19  &  189  &  17.906  &  18.041  &  18.150  &  18.409  &  18.171 \\
8  & Apr 12  &  569.1711  &  1.31  &  192  &  17.872  &  18.028  &  18.213  &  18.337  &  18.165 \\
9  & Apr 22  &  578.9261  &  1.02  &  192  &  17.870  &  18.015  &  18.166  &  18.353  &  18.164 \\
10 & Apr 24  &  580.9120  &  1.32  &  202  &  17.864  &  18.002  &  18.187  &  18.364  &  18.163 \\
11 & May 1   &  587.9478  &  1.03  &  212  &  17.847  &  18.003  &  18.190  &  18.351  &  18.161 \\
12 & May 3   &  589.9377  &  1.22  &  210  &  17.877  &  17.989  &  18.173  &  18.364  &  18.170 \\
13 & May 6   &  592.8934  &  1.19  &  212  &  17.841  &  18.039  &  18.182  &  18.369  &  18.165 \\
14 & May 11  &  597.8990  &  1.27  &  192  &  17.855  &  18.013  &  18.207  &  18.369  &  18.163 \\
15 & May 12  &  598.9280  &  0.84  &  214  &  17.836  &  18.037  &  18.175  &  18.375  &  18.160 \\
16 & May 13  &  599.9053  &  1.04  &  167  &  17.829  &  18.022  &  18.210  &  18.374  &  18.169 \\
17 & May 20  &  606.9064  &  0.90  &  163  &  17.848  &  18.031  &  18.208  &  18.345  &  18.156 \\
18 & May 21  &  607.9159  &  1.09  &  182  &  17.864  &  18.014  &  18.214  &  18.373  &  18.173 \\
19 & May 23  &  609.9184  &  1.33  &  205  &  17.857  &  18.042  &  18.205  &  18.327  &  18.161 \\
20 & May 26  &  612.9050  &  1.15  &  220  &  17.865  &  18.059  &  18.202  &  18.346  &  18.165 \\
21 & May 27  &  613.9510  &  1.15  &  231  &  17.853  &  18.044  &  18.181  &  18.381  &  18.164 \\
22 & May 28  &  614.9155  &  0.87  &  238  &  17.859  &  18.059  &  18.206  &  18.352  &  18.169 \\
23 & May 29  &  616.0342  &  1.35  &  161  &  17.848  &  18.050  &  18.215  &  18.379  &  18.171 \\
24 & Jun 5   &  622.9486  &  1.19  &  238  &  17.875  &  18.073  &  18.212  &  18.359  &  18.172 \\
25 & Jun 9   &  626.9143  &  1.36  &  206  &  17.918  &  18.031  &  18.229  &  18.344  &  18.168 \\
26 & Jun 11  &  628.9096  &  0.84  &  196  &  17.878  &  18.041  &  18.239  &  18.363  &  18.175 \\
27 & Jun 19  &  636.9364  &  0.90  &  188  &  17.887  &  18.094  &  18.266  &  18.368  &  18.176 \\
28 & Jun 23  &  640.9468  &  1.24  &  228  &  17.885  &  18.069  &  18.248  &  18.380  &  18.174 \\
29 & Jun 25  &  642.9421  &  1.29  &  207  &  17.922  &  18.072  &  18.256  &  18.335  &  18.160 \\
30 & Jul 5   &  652.9233  &  1.17  &  232  &  17.932  &  18.076  &  18.273  &  18.401  &  18.176 \\
31 & Jul 13  &  660.9244  &  1.06  &  157  &  17.950  &  18.098  &  18.302  &  18.398  &  18.178 \\
32 & Jul 21  &  668.9129  &  1.29  &  164  &  17.971  &  18.067  &  18.296  &  18.403  &  18.157 \\
33 & Jul 27  &  674.9026  &  1.27  &  202  &  17.915  &  18.120  &  18.288  &  18.451  &  18.176 \\
\enddata
\tablenotetext{a}{All frames were taken in 2008.}
\tablenotetext{b}{FWHM of the seeing disc in arcsec.}
\tablenotetext{c}{SNR of the S40 field star within a circle of radius FWHM.}
\tablenotetext{d}{$r$--SDSS brightness of A in mag. The typical error is 0.010 mag.}
\tablenotetext{e}{$r$--SDSS brightness of B in mag. The typical error is 0.012 mag.}
\tablenotetext{f}{$r$--SDSS brightness of C in mag. The typical error is 0.018 mag.}
\tablenotetext{g}{$r$--SDSS brightness of D in mag. The typical error is 0.018 mag.}
\tablenotetext{h}{$r$--SDSS brightness of S40 in mag. The typical error is 0.006 mag.}
\end{deluxetable}

\clearpage

\begin{deluxetable}{ccccc}
\tablecaption{Time delays (in days) of H1413+117.\label{tbl2}}
\tablewidth{0pt}
\tablehead{
\colhead{Method} & \colhead{Simulations} & \colhead{$\Delta \tau_{AB}$} & 
\colhead{$\Delta \tau_{AC}$} & \colhead{$\Delta \tau_{AD}$}  
}
\startdata
$D^2$    & NORMAL    & $-$17$^{+3}_{-5}$ & $-$19 $\pm$ 6       & 24$^{+5}_{-6}$\\           
         & BOOTSTRAP & $-$18 $\pm$ 3     & $-$23$^{+5}_{-7}$   & 20 $\pm$ 5\\
$\hat{\chi}^2$ & NORMAL    & $-$17 $\pm$ 3       & $-$20 $\pm$ 4       & 23 $\pm$ 4\\
               & BOOTSTRAP & $-$22$^{+3}_{-4}$   & $-$23$^{+4}_{-7}$   & 21$^{+4}_{-3}$\\
\enddata
\tablecomments{$\Delta \tau_{ij} = \tau_j - \tau_i$, so B and C are leading, and D is trailing. 
All measurements are 68\% confidence intervals.}
\end{deluxetable}

\clearpage

\begin{deluxetable}{ccc}
\tablecaption{Modelling results.\label{tbl3}}
\tablewidth{0pt}
\tablehead{
\colhead{Parameter} & \colhead{MacLeod et al. (2009)} & \colhead{This paper}  
}
\startdata
$\chi^2$/dof     	         & 4.9/5                & 7.5/7                \\ 
$b_{G1}$                   & $0\farcs 66$         & $0\farcs 68$         \\           
$\Delta \alpha_{G1}$       & $-0\farcs 166$       & $-0\farcs 165$       \\
$\Delta \delta_{G1}$       & $0\farcs 556$        & $0\farcs 552$        \\
$e_{G1}$                   & 0.26                 & 0.28                 \\
$\theta_{e_{G1}}$          & $-36\fdg 5$          & $-37\fdg 6$          \\
$b_{G2}$                   & $0\farcs 63$         & $0\farcs 45$         \\
$\Delta \alpha_{G2}$       & $\equiv -1\farcs 87$ & $\equiv -1\farcs 87$ \\
$\Delta \delta_{G2}$       & $\equiv 4\farcs 14$  & $\equiv 4\farcs 14$  \\
$\gamma_{ext}$             & 0.087                & 0.11                 \\
$\theta_{\gamma_{ext}}$    & $50\fdg 1$           & $45\fdg 4$           \\
$z_l$                      & --                   & 1.88                 \\
\enddata
\tablecomments{Position angles (ellipticity of G1 and external shear) are measured east of north and 
positions are relative to image A (negative $\Delta \alpha$ values are eastward of image A). Here, 
$b$, $e$, $\gamma$, and $z_l$ denote mass scale, ellipticity, shear strength, and lens redshift, 
respectively.}
\end{deluxetable}

\end{document}